\documentclass[aip,reprint]{revtex4-1}

\usepackage{graphicx}
\usepackage{dcolumn}
\usepackage{bm}
\usepackage[mathlines]{lineno}

\usepackage{siunitx}
\usepackage{cleveref}

\usepackage{makecell}

\draft 

\begin{document}


\title{A Low-Complexity 200 Mbit/s Full-Duplex Digital Link Over a Single Coaxial Cable for Laboratory Instrumentation}



\author{M.Wiebusch}
\email[]{m.wiebusch@gsi.de}
\affiliation{GSI Helmholtzzentrum für Schwerionenforschung (GmbH), Planckstr. 1, 64291 Darmstadt, Germany}


\date{\today}

\begin{abstract}
Modern full-duplex communication systems, such as Gigabit Ethernet, enable high data rates over challenging transmission media but require sophisticated circuitry and non-trivial digital interfaces.
For laboratory instrumentation, the objective may instead be to achieve moderate-bandwidth full-duplex communication over short, well-controlled coaxial cables with minimal circuit and interface complexity.

We present a low-complexity bidirectional digital interconnect that enables simultaneous transmission and reception of baseband logic signals over a single coaxial cable. The circuit combines a discrete passive resistive hybrid providing line termination and directional separation, a single logic gate as line driver, and a commercial differential receiver used as a comparator. No modulation, transformer coupling, adaptive calibration, or active analog or digital echo cancellation is required.
The design target is full-duplex operation at \SI{200}{\mega\bit\per\second} over \SI{10}{\meter} of standard RG-178 cable.

An analytical treatment of the hybrid network determines the system parameter that maximizes the differential signal amplitude at the receiver, while circuit simulations predict the deterministic timing errors resulting from imperfect directional separation.

Experimental measurements confirm the predicted deterministic jitter and show good agreement with the simulation results.
For typical laboratory coaxial cables up to \SI{10}{\meter}, the measured peak-to-peak edge timing error remains below \SI{1.2}{\nano\second}. A bidirectional transmission experiment with randomized data demonstrates a clearly open eye diagram. 
It was found that the transceiver inherently provides loopback functionality when the transmission line is left open.

The proposed approach provides a simple, protocol-independent bidirectional physical link which may be useful in laboratory instrumentation where cable routing or feedthrough density is constrained.
\end{abstract}


\maketitle 

\section{Introduction}

Laboratory instrumentation frequently requires the transmission of timing, trigger, synchronization, and control signals between distributed electronic systems. In many applications, these signals are carried over coaxial cables because of their well-defined impedance, excellent shielding, and widespread availability. 

When bidirectional communication is required, separate cables are typically used for transmit and receive signals, increasing cable count and connector complexity. Reducing the number of physical connections while retaining simultaneous bidirectional communication can therefore simplify experimental installations and reduce infrastructure requirements.

Examples include vacuum feedthroughs in detector systems and using pre-existing coaxial cable plants connecting experimental set-ups with data acquisition and control systems in separate areas.

This work presents a low-complexity transceiver that enables deterministic full-duplex transmission of baseband digital signals over a single coaxial cable. The proposed circuit consists only of a passive resistor bridge circuit (a so-called hybrid) and two commodity ICs that can be readily integrated on a printed circuit board (PCB) to establish digital device-to-device communication.

The proposed technique is intended for laboratory and detector instrumentation, where cable lengths are typically limited to a few meters, transmission paths are point-to-point, and the impedance of cables and interconnects is well controlled. Under these conditions, a simple resistive bridge hybrid suffices to achieve directional separation.
Based on this principle the proposed transceiver provides an inexpensive means of reducing cable count and vacuum feedthrough requirements while remaining compatible with arbitrary asynchronous digital logic signals.
The targeted design performance is full-duplex operation at \SI{200}{\mega\bit\per\second} over \SI{10}{\meter} of standard RG-178 cable.

\section{relation to existing techniques}
Whenever transmission and reception share the same physical medium, the locally transmitted signal must be suppressed at the receiver in order to properly receive and process the remote signal.
This problem arises in both wired and wireless communication systems and is addressed by duplexers, which combine and separate the transmit (TX) and receive (RX) paths at the interface to the shared transmission medium.

The two most common duplexing strategies are frequency-division duplexing (FDD) and time-division duplexing (TDD). In TDD, the communication partners transmit alternately while the remaining devices listen, resulting in half-duplex operation. The duplexing function is typically realized by an SPDT switch. A common example is RS485~\cite{mackay2003practical}, whose electrical interface supports half-duplex communication over a single twisted wire pair, while the time-division protocol is defined by higher communication layers.
Another prominent example for half-duplex on a single wire pair is 
controller area network (CAN)~\cite{CAN_ISO11898-1}.

In FDD, simultaneous transmission and reception are achieved by assigning separate frequency bands to the two directions. This requires modulation onto carrier frequencies and frequency-selective duplexers such as surface acoustic wave (SAW) or bulk acoustic wave (BAW) filters used in mobile phone RF front ends~\cite{ChenSAW}. Cable-based examples include ADSL~\cite{golden2005fundamentals} and conventional DOCSIS~\cite{FDDocsis} (Data Over Cable Service Interface Specification). Although FDD provides simultaneous bidirectional communication, the available spectral bandwidth is divided between upstream and downstream channels.

A different class of duplexers called hybrids enables simultaneous bidirectional communication within the same frequency band or even in baseband.
Here, simultaneous transmit and receive signals occupy the same frequency band and the locally transmitted signal must be suppressed by exploiting signal directionality.

Among the earliest examples are transformer-based telephone hybrids, which use a multi-winding transformer and a balancing impedance to cancel the locally transmitted signal at the receiver~\cite{Sartori}.
A closely related concept is the electrical balance duplexer (EBD), which replaces the discrete hybrid transformer by a four-port hybrid junction, a passive network with similar directional cancellation.
In combination with an electronically adjustable balance network matching and tracking the line or antenna impedance, the same isolation principle is achieved in RF front ends, often in form of a fully integrated circuit~\cite{EBD_laughlin}.

Practical hybrids provide only finite isolation, leaving residual transmitter leakage at the receiver.
In addition, reflections from impedance discontinuities produce echoes that further degrade receiver performance. Consequently, modern high-speed full-duplex systems such as Gigabit Ethernet (1000BASE-T)~\cite{ieee_ethernet} and Full Duplex DOCSIS (FDX DOCSIS)~\cite{FDDocsis} combine electrical hybrids with analog self-interference cancellation followed by adaptive digital echo cancellation to suppress both direct transmitter leakage and reflected signals~\cite{TaiCheng_GigabitEcho}.

In contrast to these approaches, the hybrid used here performs directional separation solely through a fixed passive resistive network. It neither relies on inductive coupling, as in conventional telephone hybrids and transformer-based full-duplex links, nor on active analog or adaptive digital cancellation, as employed in modern Ethernet and other high-performance full-duplex systems.

The development of modern copper-based full-duplex communication systems has largely been driven by the goal of maximizing data throughput over existing communication infrastructure, such as long copper subscriber lines, twisted-pair Ethernet cabling, or broadband cable networks. To compensate for the resulting attenuation, reflections, and impedance variations, these systems employ increasingly sophisticated analog front ends, adaptive balancing networks, and digital signal processing.

The present work addresses a different design objective.
Instead of maximizing bandwidth over challenging transmission media, it seeks to realize reliable full-duplex communication at moderate data rates ($\leq\SI{200}{MBit}$) with minimal circuit and interface complexity by exploiting the well-controlled electrical characteristics of short to medium length ($\leq\SI{10}{\meter}$) laboratory coaxial cables and impedance-matched interconnects in a simple point-to-point topology.
Under these conditions, signal attenuation and reflections are sufficiently limited that a simple resistive hybrid can provide adequate directional separation of baseband logic signals.

This enables a transceiver implementation consisting of a passive resistive hybrid built from nine standard 1\% tolerance SMD resistors, a single CMOS (Complementary Metal-Oxide-Semiconductor) logic gate as line driver, and a commercially available LVDS (Low Voltage Differential Signaling) receiver, without adaptive calibration or digital echo cancellation.

The concept uses no transformers or other forms of AC coupling, requires no modulation or data scrambling, and is therefore capable of transmitting and receiving arbitrary asynchronous logic signals, including continuous logic-high and logic-low states.
The absence of galvanic isolation implies that both transceivers share a common ground reference through the coaxial cable shield. Consequently, the proposed interface is intended for electrically compact laboratory setups in which ground potential differences are negligible. In applications involving long cable runs or distributed installations, ground loops may degrade signal integrity.

The proposed transceiver combines several characteristics that are otherwise found only separately in established communication interfaces: a single coaxial cable, a data rate comparable to 100BASE-TX Ethernet, symmetric hybrid-based full-duplex communication resembling 1000BASE-T Gigabit Ethernet, and the protocol-independent LVTTL (Low Voltage Transistor-Transistor Logic) interface simplicity of RS485 and CAN transceivers, while requiring only two commodity logic ICs and a small passive resistor network.
From the perspective of the surrounding digital logic, the transceiver behaves similar to two independent unidirectional LVDS links, one for transmission and one for reception.

The discussion above is condensed in Table~\ref{tab:comparison}, which compares the proposed transceiver with representative self-contained physical-layer interfaces in terms of duplex capability, data rate, implementation complexity, power consumption, and physical transmission medium.
DOCSIS was excluded from the comparison because it is an integrated cable-access system architecture rather than a generic physical-layer interface for point-to-point links.

\begin{table*}
\caption{\label{tab:comparison}
Comparison of representative wired communication technologies with the proposed transceiver. The comparison is restricted to self-contained physical-layer interfaces that can be implemented directly on a printed circuit board to provide digital board-to-board or device-to-device communication.
}

\begin{ruledtabular}
\begin{tabular}{ccccccc}
 \makecell{\textbf{technology}\\(example IC)} &
 \textbf{duplex} &
 \textbf{data rate} & 
 \textbf{components}\footnote{excluding connectors and decoupling capacitors} &
 \textbf{data interface} &
 \textbf{power consumption}
    \footnote{at 100\% full duplex operation or 100\% transmit if half-duplex}
    \footnote{ICs powered from VCC=\SI{3.3}{V}, except where stated otherwise}
    &
 \textbf{medium}  \\ 
 \hline
 
 \makecell{  RS485\\(SN65HVD10\cite{TI_SN65HVD10_EP})}   & 
 \makecell{  half } & 
 \makecell{  0-$\SI{25}{MBit/s}$} & 
 \makecell{  transceiver IC,\\term. resistor} &
 \makecell{  RO,$\overline{\mathrm{RE}}$,DE,DI\\(4$\times$LVTTL)} &
 \makecell{  $\gtrsim\SI{90}{mW}$\footnote{
 Power consumption is not stated in datasheet, but estimated from current required to drive a specification compliant signal: 
 $\pm\SI{1.5}{V}$ into $R_{term}||R_{term}$}  
 } &

 \makecell{  $1\times\SI{120}{\ohm}$\\ twisted pair  } 
 \\ \hline
 
 \makecell{  CAN\\(TJA1051\cite{NXP_TJA1051})}   & 
 \makecell{  half } & 
 \makecell{  0-$\SI{5}{MBit/s}$  } & 
 \makecell{  transceiver IC,\\term. resistor  } &
 \makecell{  RX, TX, EN, S\\(4$\times$LVTTL) } &
 \makecell{  $\approx\SI{250}{mW}$\\(\SI{50}{mA}, VCC=\SI{5}{\volt})  } &
 \makecell{  $1\times\SI{120}{\ohm}$\\ twisted pair  } 
 \\ \hline
 
 \makecell{  100BASE-TX\\(RTL8201\cite{RealtekRTL8201})  }   & 
 \makecell{  full,\\ separate\\lines } & 
 \makecell{  $\SI{100}{MBit/s}$} & 
 \makecell{  PHY IC,\\ magnetics,\\crystal} &
 \makecell{  MII } &
 \makecell{  $\approx\SI{300}{mW}$  } &
 \makecell{  2$\times\SI{100}{\ohm}$\\ twisted pair  } 
 \\ \hline
 
 \makecell{  1000BASE-T\\(KSZ9021\cite{MicrochipKSZ9021})  }   & 
 \makecell{  full } & 
 \makecell{  $\SI{1000}{MBit/s}$} & 
 \makecell{  PHY IC,\\ magnetics,\\crystal,\\ regulator for $\mathrm{V_{core}}$} &
 \makecell{  GMII, RGMII } &
 \makecell{  $\approx\SI{1100}{mW}$  } &
 \makecell{  4$\times\SI{100}{\ohm}$\\ twisted pair  } 
 \\ \hline
 
 \makecell{  LVDS\\(SN65LVDS1\cite{TI_SN65LVDS2},\\SN65LVDT2\cite{TI_SN65LVDS2}) }   & 
 \makecell{  full,\\ separate\\lines } & 
 \makecell{  0-$\SI{400}{MBit/s}$} & 
 \makecell{  2$\times$SOT23-5 ICs} &
 \makecell{  DIN,DOUT\\(2$\times$LVTTL) } &
 \makecell{  $\approx\SI{85}{mW}$  } &
 \makecell{  $2\times\SI{100}{\ohm}$\\twisted pair  } 
 \\ \hline
 
 \makecell{  \textbf{this work}  }   & 
 \makecell{  full } & 
 \makecell{  0-$\SI{200}{MBit/s}$} & 
 \makecell{  2$\times$SOT23-5 ICs,
 \\9 SMD resistors } &
 \makecell{  DIN,DOUT\\(2$\times$LVTTL) } &
 \makecell{  $\approx\SI{180}{mW}$  } &
 \makecell{  $1\times\SI{50}{\ohm}$\\ coax  }

\end{tabular}
\end{ruledtabular}
\end{table*}

\section{Hybrid Circuit Principle}

\begin{figure}
    \centering
    \includegraphics[width=\linewidth]{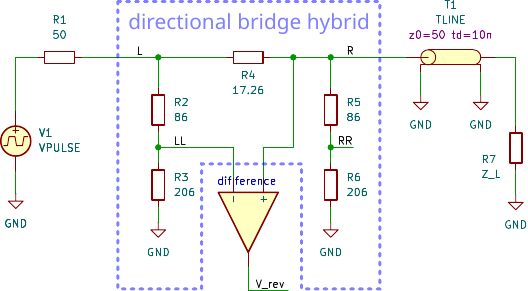}
    \caption{Circuit example of a directional bridge hybrid matched to a $\SI{50}{\ohm}$ transmission line. The output of the difference amplifier outputs only the reverse/reflected component from the far side of the transmission line.}
    \label{fig:3db_coupler}
\end{figure}

In the following we assume a basic understanding of transmission lines, impedance matching, and the need for proper termination to avoid signal artifacts when using transmission lines for communication.

The duplexing technique used in this work is a passive resistive hybrid in the form of a directional bridge, which is depicted in \cref{fig:3db_coupler}.
The presented circuit is based on a directional bridge topology described by Nash and Brunner \cite{nash2018integrated}, which was developed for broadband bidirectional RF power and return-loss measurements and which is integrated into the commercial ADL5920 \cite{AnalogDevicesADL5920}.
This work adapts the same directional-bridge principle for bidirectional digital communication.

$V1$ and $R1$ represent a signal source with \SI{50}{\ohm} output impedance driving a voltage signal into the bridge which resembles a \SI{-3}{\decibel} symmetric Pi attenuator formed by $R2$-$R6$ (marked in blue).
The bridge in turn feeds the attenuated signal into a \SI{50}{\ohm} transmission line $T1$ which is loaded with an arbitrary AC impedance $Z_L$ at the far end.
The Pi attenuator is designed to match the line impedance of \SI{50}{\ohm} and therefore acts as a resistive termination.
Standard attenuator synthesis formulas were used to determine the required resistor values \cite{vizmuller1995rf}.

Any DC or AC voltage signal applied to the left terminal $L$ of the bridge will result in an attenuated signal at the right terminal $R$.
For symmetry reasons, any voltage signal applied to the right terminal $R$ will be attenuated by the same amount in the other direction.
\begin{eqnarray}
\textrm{atten. gain:}\quad & g = & 0.71\,\,(\SI{-3}{\decibel})\\
\textrm{signal L$\rightarrow$R}:\quad & V_R = & g\cdot V_L \label{send_gain}\\
\textrm{signal L$\leftarrow$R}:\quad & V_L = & g\cdot V_R
\end{eqnarray}
The legs of the Pi attenuator are subdivided into two resistors, each forming a voltage divider.
The resistor ratio is chosen such that the voltage divider also has a voltage division factor of $g=0.71$ ($\SI{-3}{dB}$).
Consequently the voltages at the divider midpoint terminals $LL$ and $RR$ are:
\begin{eqnarray}
V_{LL} = g\cdot V_L\\
V_{RR} = g\cdot V_R.
\end{eqnarray}

We probe the voltage difference $V_{rev}$ between the right terminal $R$ and the left divider terminal $LL$.
In \cref{fig:3db_coupler} the probe is depicted as an ideal voltage difference amplifier with unity gain.
For all signals coming from the voltage source on the left side $V_{rev}$ is zero, because both terminals, $R$ and $LL$, see $V_L$ attenuated by the same factor.

\begin{eqnarray}
\textrm{signal L$\rightarrow$R}:\quad V_{rev} & = & V_R - V_{LL}\\
& = & g\cdot V_L - g\cdot V_L\\
& = & 0
\end{eqnarray}

For signals arriving from the far end of the transmission line, including those produced by reflections at the load impedance $Z_L$,
$V_{rev}$ becomes non-zero:

\begin{eqnarray}
\textrm{signal L$\leftarrow$R}:\quad V_{rev} & = & V_R - V_{LL}\\
& = & V_R - g\cdot V_L\\
& = & V_R - g\cdot g\cdot V_R\\
& = & (1-g^2)\cdot V_R\label{rec_gain}\\
& \approx & 0.5\cdot V_R\quad(\SI{-6}{\decibel})
\end{eqnarray}

The voltage $V_{rev}$ can therefore be interpreted as an attenuated version of the backward-traveling wave on the transmission line, isolated from the forward-traveling wave launched by the voltage source.
The above derivation assumes ideal resistor values and perfect impedance matching.
Owing to the symmetry of the bridge, probing the voltage difference between $L$ and $RR$ instead yields the forward-traveling wave while suppressing the reverse component.

In the present work, the unique properties of this relatively simple circuit are applied to achieve simultaneous bidirectional baseband digital communication.
Interpreting the voltage source and differential probe as the transmitter and receiver, respectively, reveals that the resistive bridge performs the essential function of a hybrid duplexer by separating forward- and reverse-traveling waves.
Unlike precision RF measurements, this application does not require perfect analog isolation. Residual leakage of the locally transmitted signal modifies the instantaneous waveform amplitude but does not necessarily lead to symbol errors, provided that no additional logic threshold crossings are introduced.
Instead, the residual coupling primarily manifests itself as a deterministic shift of the signal edge timing, which is analyzed in \cref{sec:SPICE}.

It shall be noted that the general principle is not limited to \SI{50}{\ohm} technology and the particular attenuation value of \SI{-3}{\decibel}.
The attenuator bridge shown here only illustrates the principle of directional cancellation. The actual transceiver implementation uses a different attenuation value optimized for digital threshold detection.

\section{Circuit Implementation}

\begin{figure}
    \centering
    \includegraphics[width=1\linewidth]{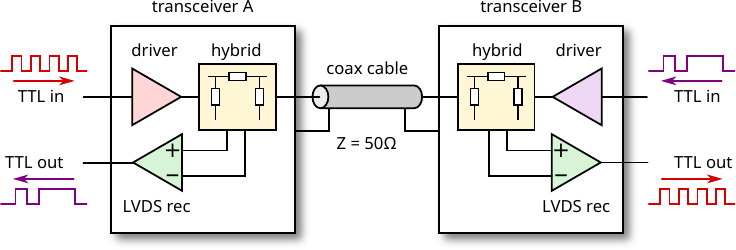}
    \caption{Conceptual diagram of two identical transceivers (comprising driver, hybrid and receiver) connected by a single $\SI{50}{\ohm}$ cable, forming a full-duplex link.
    }
    \label{fig:transceiver_AB}
\end{figure}

The intended application is illustrated in \cref{fig:transceiver_AB}. Two identical transceivers are interconnected by a single \SI{50}{\ohm} coaxial cable, forming a symmetric full-duplex point-to-point link.
Each transceiver consists of a resistive hybrid, as discussed in the previous section, and a dedicated driver and receiver IC.
A digital logic signal applied to the $TTL\_IN$ input of one transceiver is recovered at the $TTL\_OUT$ output of the remote transceiver, while communication in the opposite direction takes place simultaneously over the same cable.
The two transceivers are functionally identical and may therefore be used interchangeably at either end of the link.

\begin{figure*}
    \centering
    \includegraphics[width=0.6\linewidth]{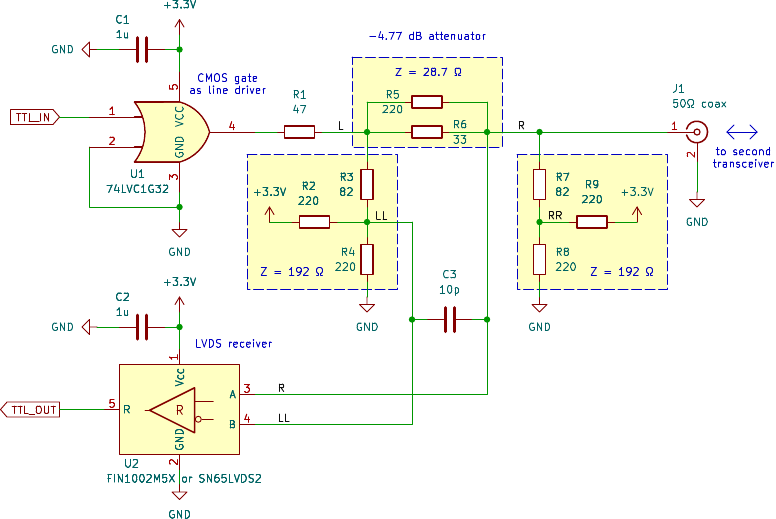}
    \caption{Circuit diagram of one bidirectional transceiver. A CMOS logic gate (U1) drives the coaxial line through a resistive hybrid which simultaneously performs directional separation and DC biasing for the LVDS receiver (U2).
    }
    \label{fig:twowaycircuit}
\end{figure*}

The circuit implementation of one transceiver is shown in \cref{fig:twowaycircuit}. Although more elaborate than the conceptual bridge presented in \cref{fig:3db_coupler}, it retains the same three functional building blocks:

\begin{itemize}
\item A line driver ($U1$: 74LVC1G32), implemented as a single CMOS OR gate configured as a non-inverting LVTTL logic buffer. The driver takes the place of the voltage source.

\item An LVDS receiver ($U2$: FIN1002M5X~\cite{fin1002_datasheet} or SN65LVDS2~\cite{TI_SN65LVDS2}) without integrated termination, replacing the ideal differential probe introduced in the previous section. The receiver is operated as a high-speed differential comparator.

\item A resistive bridge hybrid derived from a Pi attenuator with integrated voltage-divider legs.
In contrast to the conceptual circuit, the lower branch of each divider is split into two equal resistors connected to GND and VCC, respectively.
\end{itemize}

The transceiver operates from a single \SI{3.3}{\volt} supply. Unlike the conceptual bridge discussed in the previous section, which considers arbitrary voltage signals, the implemented circuit interfaces directly to LVTTL logic levels and therefore drives discrete voltage levels onto the transmission line.

The dashed yellow boxes in \cref{fig:twowaycircuit} identify the three resistor groups that functionally correspond to the elements of a Pi attenuator with an attenuation factor of $g=0.577$ (\SI{-4.77}{\decibel}) and a characteristic impedance of $Z_{in}=Z_{out}=\SI{50}{\ohm}$. The \emph{Th\'evenin} equivalent resistance of the upper resistor group is \SI{28.7}{\ohm}, while each attenuator leg exhibits an equivalent resistance of \SI{192}{\ohm}. The two legs simultaneously form voltage dividers with the same division factor of $g=0.577$.

The implemented resistor values do not exactly match the theoretical values (\SI{28.86}{\ohm} and \SI{186.7}{\ohm}), but are rounded to the nearest preferred values of the E12 series. The upper resistor is realized by the parallel combination of $R5$ and $R6$, allowing the desired resistance to be approximated using readily available components.

The additional resistors connected to \SI{3.3}{\volt} ($R2$ and $R9$) shift the operating point of the otherwise linear resistive network from ground to $VCC/2$. As a result, all attenuation and voltage division takes place with respect to $VCC/2$ rather than $GND$.
This DC level shift serves two purposes: it establishes a well-defined positive or negative differential input voltage for every possible transmitter state, symmetric about \SI{0}{\volt}, while simultaneously biasing both receiver inputs within the specified LVDS common-mode input voltage range~\cite{diodes_LVDS}.

A small capacitor ($C3=\SI{10}{\pico\farad}$) is connected across the differential receiver inputs to suppress high-frequency transients.

Compared with the illustrative \SI{-3}{\decibel} bridge discussed previously, the attenuation has been increased to \SI{-4.77}{\decibel} to maximize the differential signal amplitude at the receiver while maintaining impedance matching and circuit symmetry. The optimum attenuation can be obtained by solving the following extremum problem:

The overall voltage gain from the driver output of transceiver A to the differential receiver input of transceiver B is the product of three factors.
We approximate the CMOS driver as an ideal voltage source switching between \SI{0}{\volt} and $VCC$. Due to $R1\approx\SI{50}{\ohm}$ in series with the driver and the attenuator matched to $\SI{50}{\ohm}$, one half of the driver's voltage appears across its series resistor, while the remaining half is applied to the attenuator input.
The Pi attenuator contributes a factor of $g$, whereas the receiver bridge contributes a factor of $(1-g^2)$, as derived in (\ref{rec_gain}). The resulting end-to-end gain is therefore

\begin{eqnarray}
G_{\mathrm{total}}(g)=\frac{1}{2}\,g\,(1-g^2),
\end{eqnarray}

which is to be maximized as a function of attenuation factor $g$.

The optimum attenuation is readily obtained by differentiating the end-to-end gain with respect to $g$ and equating the first derivative to zero:

\begin{eqnarray}
\frac{d G_{\mathrm{total}}(g)}{dg} & = & \frac{1}{2}-\frac{3}{2}g^2 = 0\\
\Rightarrow g & = & \sqrt{\frac{1}{3}} \approx 0.577\\
\Rightarrow G_{\mathrm{total}} & = & \frac{\sqrt{3}}{9} \approx 0.192.
\end{eqnarray}

For a driver supply voltage of \SI{3.3}{\volt}, this corresponds to a differential receiver input swing of
$\SI{3.3}{\volt}\cdot0.192=\SI{0.64}{\volt}$, or $\pm\SI{317}{\milli\volt}$ about the common-mode voltage. This value closely matches the nominal differential output swing ($\pm\SI{350}{\milli\volt}$) of a standard LVDS driver~\cite{diodes_LVDS}, indicating that the proposed passive bridge produces signal levels well suited to direct LVDS reception.

\section{SPICE Simulation}

\label{sec:SPICE}

\begin{figure*}
    \centering
    \includegraphics[width=0.7\linewidth]{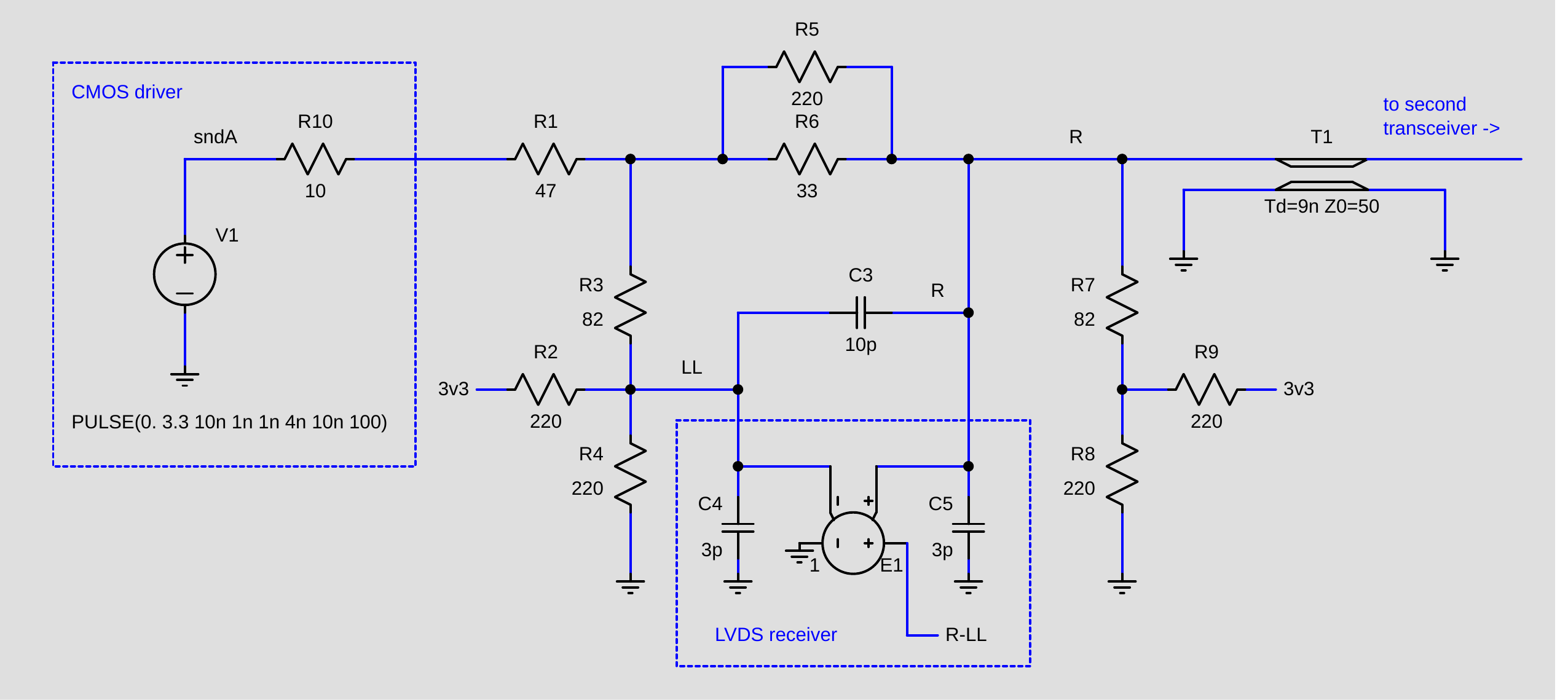}
    \caption{SPICE model of one transceiver used for simulation. The lossless $\SI{50}{\ohm}$ transmission line connects to an identical remote transceiver (not shown).}
    \label{fig:spice_schematic}
\end{figure*}

To illustrate the dynamics and resulting voltage levels within the hybrid network, a SPICE (Simulation Program with Integrated Circuit Emphasis) simulation is performed. Two identical transceivers are connected by an ideal transmission line, analogous to the configuration shown in \cref{fig:transceiver_AB}.
The simulation serves two purposes: first, to verify the expected differential signal levels derived in the previous section, and second, to investigate the residual interaction between simultaneous transmit and receive signals.

The SPICE model of transceiver A is shown in \cref{fig:spice_schematic}. The CMOS line driver is represented by a pulsed voltage source with a rise time of \SI{1}{\nano\second}, approximating the typical output slope of the 74LVC1G32 under moderate capacitive loading. An additional series resistance of \SI{10}{\ohm} models the driver's finite output impedance.

The initial objective of the simulation is to determine the differential voltage presented to the LVDS receiver. 
Accordingly, the receiver is initially modeled as an ideal differential voltage probe (voltage-controlled voltage source). Comparator dynamics and internal hysteresis are neglected at this stage and are incorporated later in the simulation.
To account for the receiver input capacitance, capacitors $C4$ and $C5$ of \SI{3}{\pico\farad} are included, corresponding to the specified input capacitance of \SI{2.3}{\pico\farad} per input pin~\cite{fin1002_datasheet} plus the estimated capacitance of the PCB pads.

For the first simulation scenario, transceiver A transmits a \SI{100}{\mega\hertz} square wave, while transceiver B simultaneously transmits a similar square wave at \SI{75}{\mega\hertz}.
The transmission line is assumed to be lossless and assigned a propagation delay of \SI{9}{\nano\second}, avoiding an accidental alignment of the propagation delay with either stimulus period.

\begin{figure}
    \centering
    \includegraphics[width=1\linewidth]{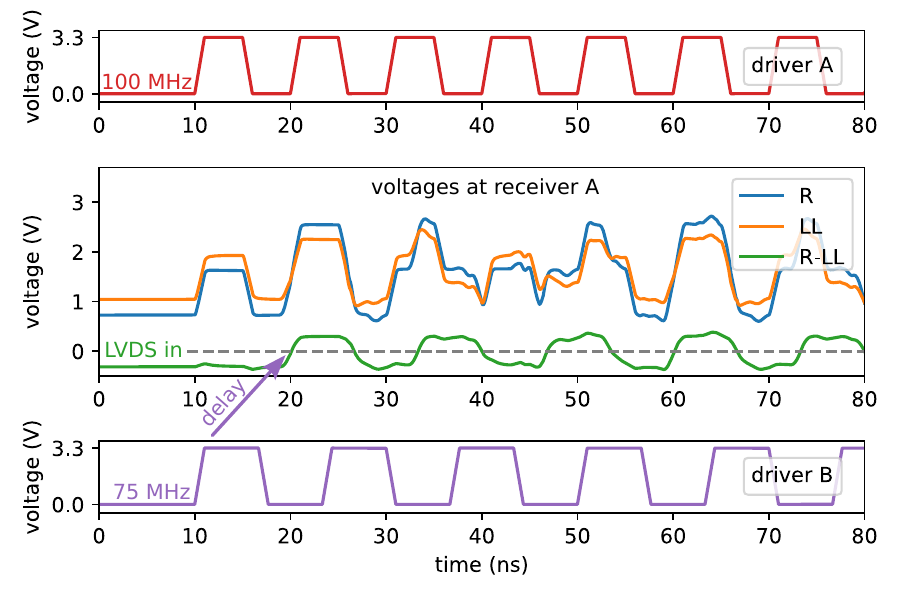}
    \caption{
    Simulated full-duplex operation of two transceivers connected by a $\SI{50}{\ohm}$ transmission line (9 ns delay). Top and bottom panels show the driver voltages of transceivers A (100 MHz) and B (75 MHz); the middle panel displays node voltages at the differential inputs of receiver A.}
    \label{fig:spice_waveform}
\end{figure}

The simulated waveforms are shown in \cref{fig:spice_waveform}.
The outer panels display the driver output voltages of the two transceivers, while the center panel shows the voltages at nodes $R$ and $LL$ of transceiver A, corresponding to the two inputs of the LVDS receiver.
Owing to the biasing network, both input voltages remain within the specified LVDS common-mode input voltage range of \SIrange{0.5}{2.5}{\volt}.
Their differential voltage forms an LVDS-compatible waveform centered about \SI{0}{\volt}, closely reproducing the signal transmitted by transceiver B, albeit with noticeable waveform distortion.

\begin{figure}
    \centering
    \includegraphics[width=1\linewidth]{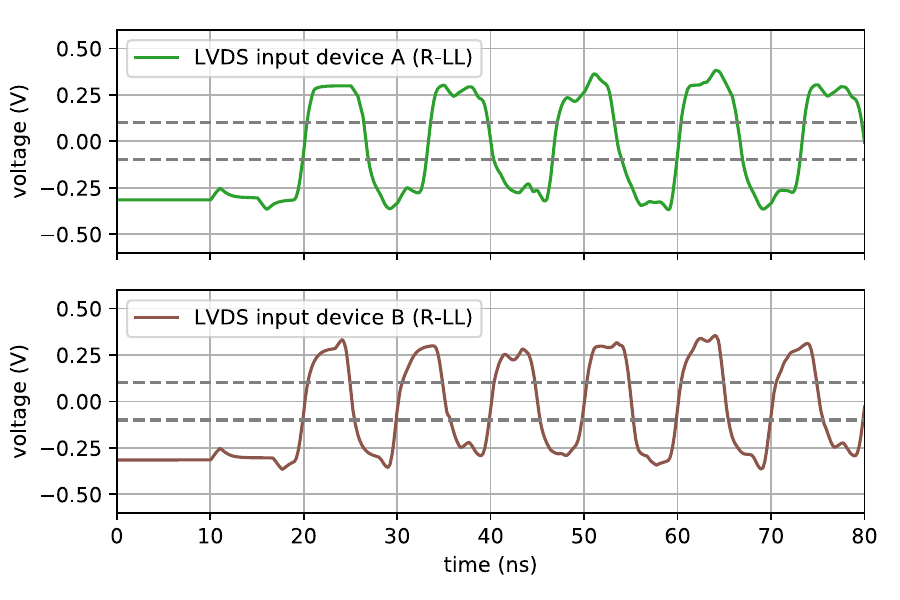}
    \caption{
    Differential input voltages (R–LL) at the LVDS receivers of transceivers A and B under simultaneous transmission. Dashed gray lines indicate the maximum guaranteed switching thresholds ($\pm\SI{100}{\milli\volt}$) of a standard LVDS receiver.}
    \label{fig:spice_wfm_RLL}
\end{figure}

A magnified view of the differential receiver waveforms for both transceivers is shown in \cref{fig:spice_wfm_RLL}.
In the ideal case of a perfectly matched resistive hybrid, ideal resistor values, and negligible receiver input capacitance, the transmitted and received signals would be perfectly separated, resulting in an undistorted differential receive waveform.

To obtain a more realistic model, the simulation includes resistor mismatches of up to approximately \SI{3}{\percent} arising from the use of E12 resistor values, an intentional deviation of the driver output impedance from its nominal value, and the finite input capacitance of the LVDS receiver. As a consequence, a small fraction of the locally transmitted signal leaks into the receiver path, while the imperfect impedance match gives rise to weak reflections on the transmission line.
These effects intentionally approximate the non-ideal behavior of a practical implementation constructed from standard-value and standard-tolerance components.

In the present simulation, the dominant source of waveform distortion is the finite input capacitance of the LVDS receiver, which prevents the receiver from sensing the exact voltage difference between the corresponding hybrid nodes.
Owing to the resistive bridge topology, the two receiver inputs exhibit different effective RC time constants, as they are driven through different source impedances despite having nearly identical input capacitances.
Consequently, one input responds slightly faster than the other, producing spikes and depressions in the differential waveform during signal transitions.
The capacitor $C3=\SI{10}{\pico F}$ connected across the receiver inputs attenuates these transient artifacts at the expense of a modest reduction in bandwidth.

Despite these non-idealities, the simulated differential waveform remains well suited for reliable digital communication. The zero crossings retain sufficiently steep edges for precise timing, while the differential amplitude of approximately $\pm\SI{250}{\milli\volt}$ corresponds to the minimum LVDS-compliant output swing and remains well above the maximum guaranteed LVDS receiver threshold of $\pm\SI{100}{\milli\volt}$~\cite{analog_LVDS}.
Empirical measurements indicate that the actual switching threshold of the FIN1002M5X LVDS receiver is considerably lower than the guaranteed datasheet value.

Nevertheless, imperfect directional separation causes the locally transmitted signal to perturb the received waveform, resulting in a deterministic edge timing error during simultaneous full-duplex operation. To first order, the resulting timing error $\Delta t$ is proportional to the instantaneous amplitude perturbation $\Delta V$ divided by the signal edge slope at the threshold crossing\cite{Sepke2009}:

\begin{eqnarray}
\Delta t \approx \frac{\Delta V}{dV/dt}\label{eqn:slopejitter}
\end{eqnarray}

To quantify this effect, a second simulation scenario is considered. As before, transceiver A transmits a \SI{100}{\mega\hertz} square wave ($T=\SI{10}{\nano\second}$), while transceiver B transmits a similar waveform with a period of $(10+\frac{\pi}{10})\,\si{\nano\second}$. The irrational period offset ensures that all relative phase relationships between the two transmitters are sampled over the simulation interval.
A \SI{100}{\mega\hertz} square wave was chosen as the test signal because it represents the worst-case transition density of a binary serial data stream at \SI{200}{\mega\bit\per\second}, corresponding to the target data rate of the proposed transceiver.

The simulation is performed over \SI{20}{\micro\second}. A total of 1998 rising edges transmitted by device A and the corresponding edges received by device B are evaluated by a dedicated Python script. During post-processing, the differential receiver waveform is converted into logic transitions using a comparator model with a hysteresis of \SI{20}{\milli\volt}, based on prior empirical measurements.

\begin{figure}
    \centering
    \includegraphics[width=1\linewidth]{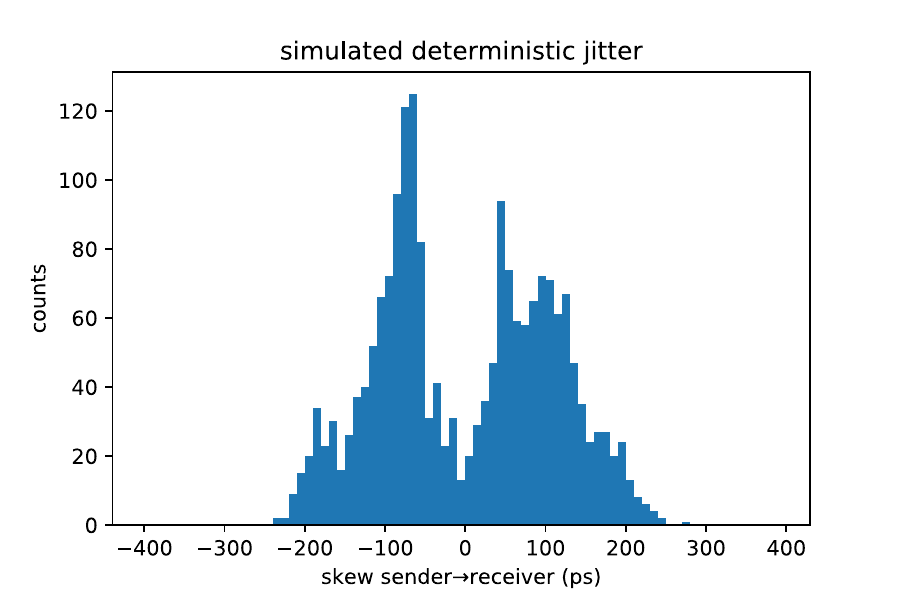}
    \caption{
    Simulated distribution of deterministic jitter due to the imperfect separation between transmitted and received channels in full-duplex mode. 
    }
    \label{fig:detjitter}
\end{figure}

The resulting distribution of edge timing errors is shown in \cref{fig:detjitter}.
However its exact shape depends strongly on the cable length and test pattern periods, indicating that reflections caused by imperfect impedance matching contribute to the deterministic jitter mechanism. The simulated peak-to-peak timing spread is approximately \SI{500}{\pico\second}. Since the simulation does not include noise sources, this timing variation represents deterministic jitter exclusively.
Peak-to-peak jitter is used throughout this work instead of a statistical measure such as the standard deviation, as the proposed transceiver operates directly on asynchronous logic signals without clock recovery or error correction. Consequently, the worst-case edge timing error determines the reliable operating margin of the link.

In conclusion, the SPICE simulations demonstrate that the proposed transceiver
\begin{enumerate}
    \item maintains the receiver input voltages within the specified LVDS common-mode input range,
    \item produces sufficient differential signal amplitude with adequate receiver threshold margin, and
    \item exhibits a deterministic timing error during simultaneous full-duplex operation in the order of \SI{500}{\pico\second} peak-to-peak, depending on the interplay between test patterns and delay line length.
\end{enumerate}

A simulated systematic sweep over a varying cable length is presented in the next section alongside experimental data.

\section{Experimental Results}

\begin{figure}
    \centering
    \includegraphics[width=1\linewidth]{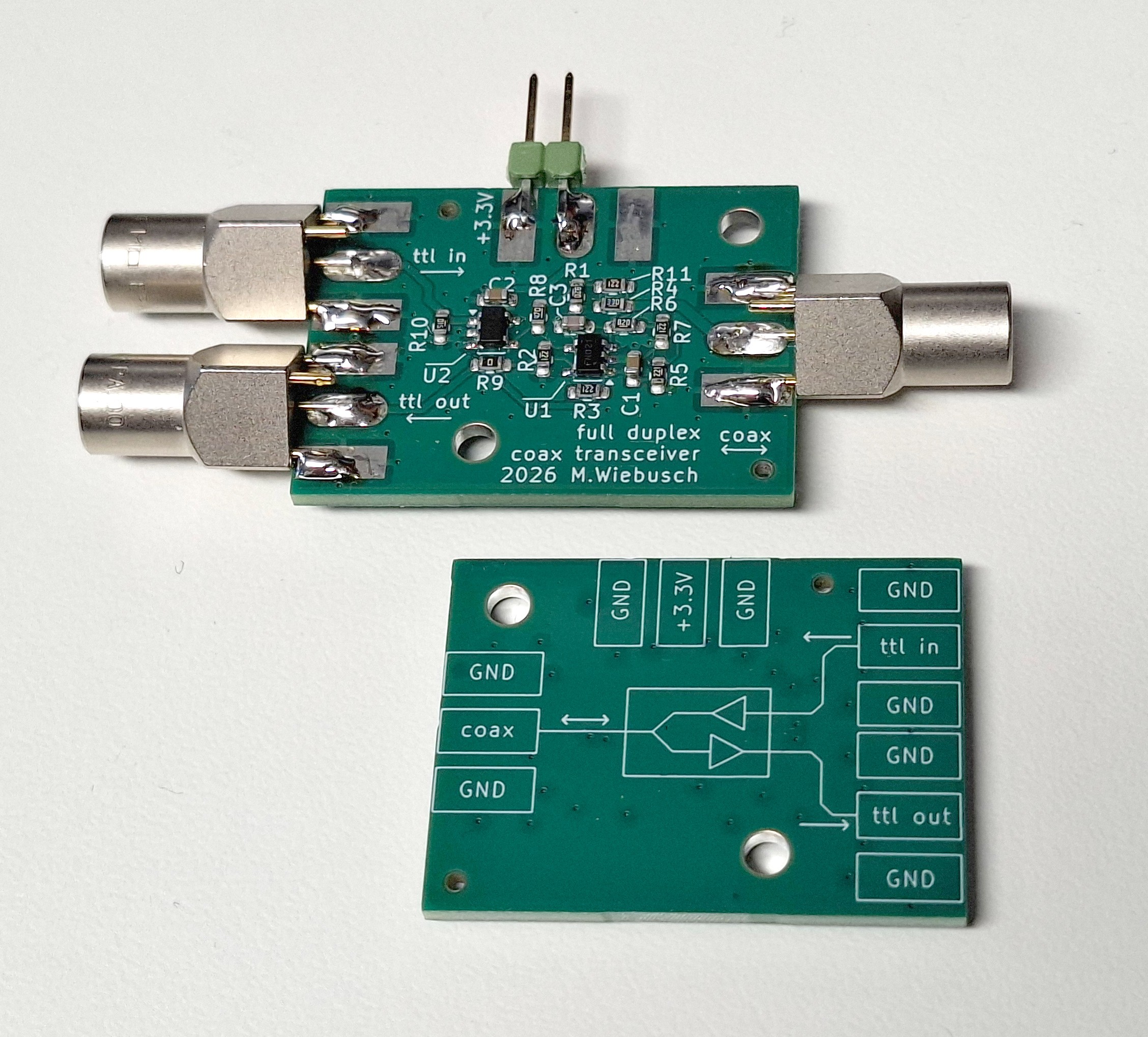}
    \caption{Photograph of one transceiver board (front/back). The board dimensions are $33\times\SI{25}{\milli\meter^2}$.
    The edge connector footprints allow for soldering either pin headers, SMA or LEMO00 (shown here).
    }
    \label{fig:board_photo}
\end{figure}


The proposed transceiver was implemented as a standalone printed circuit board measuring $33\times\SI{25}{\milli\meter}^2$. A photograph of the assembled board is shown in \cref{fig:board_photo}. Apart from an optional \SI{50}{\ohm} termination resistor connected in parallel with the LVTTL input, the implementation is identical to the circuit shown in \cref{fig:twowaycircuit}. When the transceiver is driven by a laboratory waveform generator, this resistor provides proper termination of the input cable.

Since sub-\SI{10}{\pico\second} timing performance was not anticipated, a conventional two-layer PCB stack-up was considered sufficient. Nevertheless, standard high-frequency layout practices were followed: signal traces were kept short to minimize parasitic inductance, and the bottom layer was implemented as a continuous ground plane to provide a well-defined return current path.
Each transceiver comprises two SOT23-5 integrated circuits, three 0603 capacitors, ten 0603 resistors, and the required coaxial connectors.


The transceiver was evaluated experimentally using a test setup analogous to the SPICE jitter study described in \cref{sec:SPICE}. Two identical transceivers were connected by a standard RG-178 laboratory coaxial cable with LEMO 00 connectors on both ends. The cable length was varied by combining different cable segments in increments of \SI{20}{\centi\meter} up to a total length of \SI{11}{\meter}, exceeding the design goal of \SI{10}{\meter}.

The transceiver inputs were driven by the two channels of a dual-channel arbitrary waveform generator, while a digital oscilloscope samples the transceiver outputs.
The laboratory equipment used for signal generation and measurement is listed in \cref{sec:methods}.
The waveform generator and oscilloscope were connected to the transceivers using \SI{1}{\meter}-long RG-178 coaxial cables.
One channel generated a \SI{100}{\mega\hertz} square wave applied to transceiver A, while the second generated a \SI{96.954}{\mega\hertz} square wave applied to transceiver B.
Analogous to the procedure in \cref{sec:SPICE}, the second frequency corresponds to a wave period of $T=(10+\pi/10)~\mathrm{ns}$, which has no common multiples with the period of the \SI{100}{\mega\hertz} signal.
The chosen frequencies ensured that all relative phase relationships between the two transmitted signals were sampled over time.

As discussed before, a \SI{100}{\mega\hertz} square wave was chosen as it represents the worst-case transition density of a binary serial data stream at \SI{200}{\mega\bit\per\second}.
The measured current consumption of one transceiver is \SI{55}{\milli\ampere} at \SI{3.3}{V} during \SI{100}{\mega\hertz} operation and approximately halved when transmitting a constant logic-low level.

The outputs of both transceivers were recorded with a digital oscilloscope capable of high-resolution edge timing measurements (better than \SI{10}{\pico\second}). 
The arrival times of the \SI{100}{\mega\hertz} signal at the output of transceiver B were measured relative to the waveform generator's reference clock output, allowing the deterministic timing error introduced by the full-duplex link to be quantified.

\begin{figure}
    \centering
    \includegraphics[width=1\linewidth]{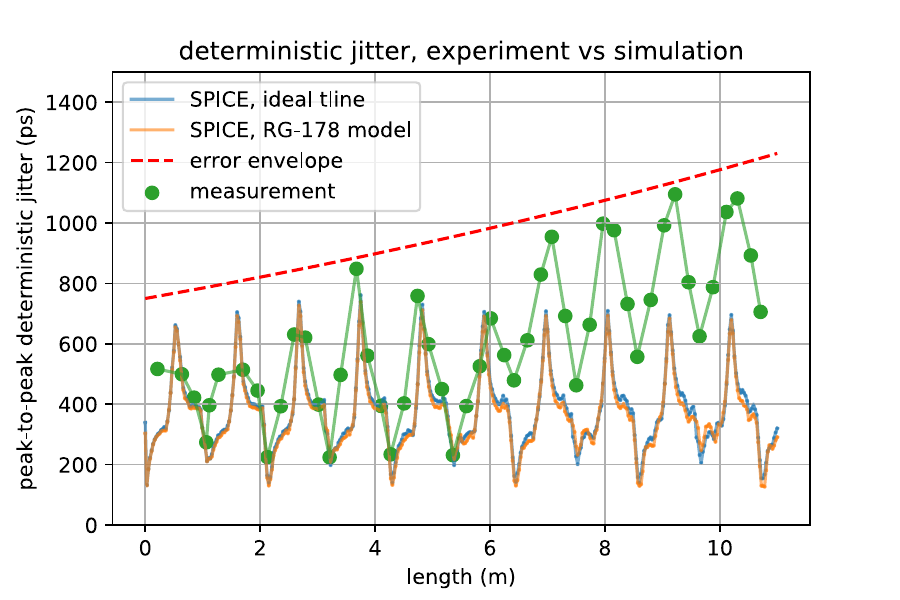}
    \caption{Measured peak-to-peak deterministic jitter as a function of cable length, compared with SPICE predictions using both ideal and lossy transmission-line models.
    }
    \label{fig:jitter_vs_length}
\end{figure}

\begin{table*}
\caption{
Comparison of measured and simulated jitter extrema for cable lengths up to \SI{6}{\meter}.
The final row in each section gives the arithmetic mean of the five extrema.
}
\label{tab:jitter_extrema}
\begin{ruledtabular}
\begin{tabular}{ccccccc}
Extremum & $n$ & Cable length (\si{\meter}) &
Measurement (\si{\pico\second}) &
Simulation (\si{\pico\second}) &
\makecell{ Deviation (\si{\pico\second})\\ (meas. - sim.) } & 
\makecell{ Deviation (\si{\percent})\\ (meas. vs sim.) } \\
\hline
Peak   & 1 & 0.52 & 503 & 662 & -159 & \SI{-24}{\percent} \\
       & 2 & 1.60 & 513 & 705 & -192 & \SI{-27}{\percent} \\
       & 3 & 2.67 & 624 & 740 & -116 & \SI{-16}{\percent} \\
       & 4 & 3.75 & 845 & 762 &   83 & +\SI{11}{\percent} \\
       & 5 & 4.83 & 757 & 728 &   28 & +\SI{4}{\percent}  \\
Average & -- & 1.08 (period) & 648 & 719 & -71 & \SI{-10}{\percent} \\
\hline
Trough & 1 & 1.07 & 273 & 211 &  62 & +\SI{29}{\percent} \\
       & 2 & 2.15 & 223 & 142 &  81 & +\SI{57}{\percent} \\
       & 3 & 3.22 & 221 & 197 &  24 & +\SI{12}{\percent} \\
       & 4 & 4.29 & 233 & 138 &  95 & +\SI{69}{\percent} \\
       & 5 & 5.38 & 230 & 199 &  31 & +\SI{15}{\percent}  \\
Average & -- & 1.08 (period) & 236 & 177 & 58 & +\SI{33}{\percent} \\
\end{tabular}
\end{ruledtabular}
\end{table*}

The measured peak-to-peak jitter as a function of cable length is shown in \cref{fig:jitter_vs_length}, alongside the corresponding values predicted by the SPICE simulation, whose continuously variable transmission line delay permits a finer sampling of simulated cable lengths.
Both simulation and measurement exhibit a pronounced periodic dependence of the peak-to-peak jitter on cable length, with a period of approximately \SI{1.08}{\meter}. The locations of the maxima and minima agree well between simulation and experiment. For cable lengths below approximately \SI{6}{\meter}, the simulated and measured timing error oscillates between roughly \SI{200}{\pico\second} and \SI{800}{\pico\second}. For longer cables an additional jitter contribution becomes apparent in the measured data that increases with cable length, and which is not captured by the simulation. At the maximum design target length of \SI{10}{\meter} the worst case jitter remains below \SI{1.2}{\nano\second}.

A quantitative comparison of the first five maxima and minima, occurring at cable lengths below approximately \SI{6}{\meter}, is provided in \cref{tab:jitter_extrema}. In this range, before the systematic increase in measured jitter becomes apparent, the absolute deviations between measurement and simulation remain below \SI{200}{\pico\second} for all maxima and below \SI{100}{\pico\second} for all minima. 

By disconnecting the second square wave generator, a unidirectional transmission scenario is realized.
In this mode, the jitter between received signal and the reference clock output remains below \SI{50}{\pico\second} peak-to-peak independent of cable length.

The differential signal amplitude at the LVDS receiver input was measured using a differential probe. After \SI{1}{\meter} of RG-178 cable, a differential amplitude of \SI{\pm 290}{\milli\volt} was measured, decreasing to \SI{\pm 220}{\milli\volt} at the maximum design target length of \SI{10}{\meter}. The corresponding signal rise time was also investigated. A 20--80\% rise time of \SI{1.1}{\nano\second} was measured after \SI{1}{\meter} of cable, increasing to \SI{1.9}{\nano\second} after \SI{10}{\meter}.

 \begin{figure}
     \centering
     \includegraphics[width=1\linewidth]{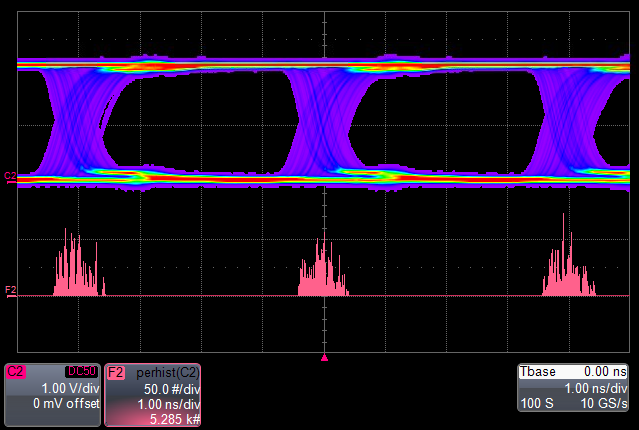}\\(a)
     \\
     \includegraphics[width=1\linewidth]{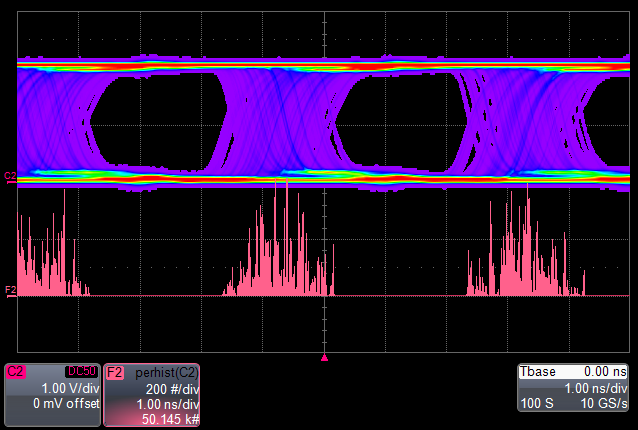}\\(b)
     \caption{Eye diagram of the receiver output signal for a bidirectional transceiver pair connected by a \SI{3.2}{\meter} (a) and a \SI{16.4}{\meter} (b) coaxial cable. Two independent randomized serial data streams at \SI{250}{\mega\bit\per\second} are transmitted simultaneously in opposite directions.}
     \label{fig:eye250}
 \end{figure}

An additional experiment was performed to evaluate the transmission of arbitrary data. Instead of periodic square waves, another arbitrary waveform generator produced two independent (pseudo-)random serial data streams at \SI{250}{\mega\bit\per\second}.
The data rate was intentionally chosen to exceed the stated design target of \SI{200}{\mega\bit\per\second} in order to demonstrate the available performance margin.
For this test, the cable length was set to \SI{3.2}{\meter}, corresponding to an empirically determined local maximum of the deterministic jitter for the new waveform.

The resulting eye diagram, shown in \cref{fig:eye250}\,(a), was recorded using the above mentioned digital oscilloscope recording the TTL output of transceiver B. The acquisition was performed continuously over several minutes. The measured peak-to-peak jitter amounts to approximately \SI{870}{\pico\second}. Despite this timing uncertainty, the eye remains clearly open throughout the measurement, indicating that the timing and voltage margin is sufficient for error-free serial data transmission under these conditions.
However, it needs to be stated that no explicit bit-error-rate measurements were performed.

To illustrate the degradation of the received waveform with increasing cable length, a second eye diagram (see \cref{fig:eye250}\,(b)) was recorded under otherwise identical conditions using a cable length of \SI{16.4}{\meter}, corresponding to another empirical maximum of the deterministic jitter. The measured peak-to-peak timing jitter increases to approximately \SI{1.9}{\nano\second} while the eye remains open.
This cable length lies beyond the intended operating range of the proposed transceiver and serves to illustrate the available design margin. 


The transceiver was also evaluated with a modified resistive hybrid adapted for \SI{75}{\ohm} transmission lines. The hybrid resistor values were scaled by a factor of 1.5 to preserve the corresponding impedance relationships and maintain proper line termination. Using a \SI{3}{\meter} television antenna cable, the measured peak-to-peak jitter was comparable to that obtained with the \SI{50}{\ohm} configuration, demonstrating that the circuit principle is not limited to \SI{50}{\ohm} transmission lines.

\section{Discussion}
The observed periodicity in the deterministic jitter is attributable to the interplay between the test signal wavelength and cable length. Owing to the imperfect line termination at the transceivers, the ensuing residual reflections interfere with subsequent signal transitions. Depending on the propagation delay, these reflections either reinforce or reduce the instantaneous waveform distortion, leading to the observed periodic modulation of the deterministic jitter. This effect is well captured by the SPICE simulation, and both the predicted periodicity and the magnitude of the effect agree well with the experimental data for cable lengths up to approximately \SI{6}{\meter}.

The experimental sampling of cable length is constrained by the available physical cable configurations, whereas the simulation permits continuous variation of the transmission-line delay. Consequently, the measured jitter curve may undersample the periodic modulation, such that the apparent locations and amplitudes of its peaks and troughs do not necessarily coincide with the actual extrema. This sampling limitation should be considered when comparing individual measurement points with the simulated curve.

The simulation is not intended as a precision predictor of every detail of the measured waveform. Nevertheless, the comparatively simple SPICE model reproduces the average jitter level and the depth of its periodic modulation reasonably well as the cable length increases. In particular, the measured and simulated curves exhibit closely aligned periodicity and remain in phase over the entire investigated length range. This agreement is obtained using general assumptions about the circuit and signal source, without fine-tuning simulation parameters to fit the experimental data. The results therefore indicate that the model captures the dominant physical mechanism governing the cable-length dependence of the deterministic jitter, even though it does not reproduce all details of the experimental response.


For cables longer than \SI{6}{\meter}, the experimental jitter values indicate an additional timing error that increases with cable length and which is not captured by the present simulation model. To investigate this discrepancy, an additional SPICE simulation was performed using a lossy transmission-line model representing RG-178 coaxial cable ($R'=\SI{0.8}{\ohm/\meter}$, $C'=\SI{96}{\pico F/\meter}$, $Z_0=\SI{50}{\ohm}$ according to~\cite{tasker_rg178bu_datasheet}, and $L'=Z_0^2C'$). The resulting simulated jitter is overlaid with the previous data in \cref{fig:jitter_vs_length}. It differs only marginally from that obtained with the ideal transmission-line model and therefore does not reproduce the experimentally observed systematic increase in timing error.
The additional timing error is therefore likely caused by edge degradation due to frequency-dependent cable losses, such as dielectric loss and skin effect, which are not fully represented by the simplified lossy transmission-line model available in the SPICE simulation. According to equation~(\ref{eqn:slopejitter}), reduced edge steepness directly increases the sensitivity to (self-)interference and therefore leads to a larger timing error.
This interpretation is consistent with the measured increase in differential signal rise time from \SI{1.1}{\nano\second} to \SI{1.9}{\nano\second} for cable lengths of \SI{1}{\meter} and \SI{10}{\meter}, as the increase in rise time is accompanied by an approximately proportional increase in the worst-case jitter.

The results obtained in the unidirectional configuration provide a useful reference for separating the intrinsic effects of electronics and transmission medium from those introduced by the self-interference due to full-duplex operation.
With the second signal source disconnected, the measured peak-to-peak jitter remains below \SI{50}{\pico\second} and shows no measurable dependence on cable length. This is approximately an order of magnitude smaller than the deterministic jitter observed in full-duplex operation, indicating that intrinsic electronic timing uncertainty, including contributions from thermal noise, is substantially smaller than the timing error caused by residual self-interference.
Thus, self-interference can be identified as the dominant source of timing uncertainty in the proposed transceiver.

The available differential signal amplitude provides further evidence that voltage swing is not a limiting factor. Even after \SI{10}{\meter} of RG-178 cable, a differential amplitude of \SI{\pm220}{\milli\volt} is present at the LVDS receiver input. This exceeds the worst-case LVDS switching threshold of \SI{\pm100}{\milli\volt} by more than a factor of two, leaving a substantial voltage margin at the maximum design target cable length.

As discussed previously, the proposed transceiver operates directly on asynchronous logic signals. Consequently, peak-to-peak jitter, i.e., the worst-case edge timing error, is used throughout this work as the primary measure of transmission quality. To estimate the performance for arbitrary logic signals and cable lengths rather than the specific test pattern and cable lengths evaluated above, it is useful to consider the envelope of the simulated and measured jitter curves (dashed red line in \cref{fig:jitter_vs_length}). Based on this envelope, the proposed transceiver is expected to preserve edge timing to better than \SI{1}{\nano\second} for typical laboratory cables up to approximately \SI{6}{\meter}, and better than \SI{1.2}{\nano\second} for cable lengths up to \SI{10}{\meter}.

The eye-diagram measurements with randomized serial data complement these deterministic jitter measurements by demonstrating the resulting timing margin for a realistic sequence of data transitions. At a cable length of \SI{3.2}{\meter}, the eye diagram measured at a symbol rate of \SI{250}{\mega\bit\per\second} exhibits an edge timing uncertainty of \SI{870}{\pico\second} which is consistent with the previously discussed jitter envelope. The achieved symbol rate exceeds the design target of \SI{200}{\mega\bit\per\second}, demonstrating that the transceiver provides a measurable performance margin relative to its target specification. The eye remains clearly open under these conditions, which does not by itself prove error-free transmission but provides a useful indication that sufficient timing and voltage margin remains for reliable symbol transmission.

A second eye diagram was recorded under an intentionally more demanding condition: a symbol rate of \SI{250}{\mega\bit\per\second} over \SI{16.4}{\meter} of cable, now also exceeding the maximum design target length of \SI{10}{\meter}. The persistence of a substantial eye opening indicates that signal amplitude is not the primary limiting factor. However the reduced edge steepness resulting from frequency dependent cable attenuation progressively reduces the available timing margin.
Under these conditions, the peak-to-peak timing uncertainty (\SI{1.9}{\nano\second}) becomes comparable to the symbol period, making reliable operation progressively more challenging.

The timing characteristics define the range of applications for which the proposed transceiver is appropriate. For applications in which picosecond-level timing precision is essential, such as the transmission of timing-critical signals from time-of-flight detectors, the proposed technology is not recommended, at least not in full-duplex operation. For less timing-critical asynchronous signals, such as readout-request triggers that can tolerate an additional nanosecond of timing uncertainty, the measured timing performance is sufficient.
The same argument applies to moderate-bandwidth digital serial links, given the receiving device manages to sample the symbol with a sufficient timing margin relative to the jitter-affected transition edge.

The measurements presented in this work were performed under controlled laboratory conditions. In a field application, additional sources of timing uncertainty and potential symbol errors may arise from insufficiently filtered power-supply ripple, electromagnetic interference, or grounding-related effects such as ground loops. These effects are not specific to the proposed transceiver and represent general caveats for high-speed digital systems deployed outside controlled laboratory environments. A comprehensive analysis of all possible fault modes and environmental conditions is beyond the scope of this work.

Based on the above discussion, the design target of \SI{200}{\mega\bit\per\second} over a cable length of \SI{10}{\meter} should therefore be regarded not as a hard performance limit, but as a conservative operating recommendation with substantial margin. This margin is expected to provide some tolerance to application-specific noise and interference in practical implementations.

From an interfacing perspective, the proposed bidirectional transceiver behaves similarly to a pair of conventional LVDS transmitter and receiver ICs using two separate lines.
However, during full-duplex operation, the transceiver consumes approximately \SI{180}{\milli\watt}, about twice the power consumption of a comparable LVDS transmitter/receiver pair (see \cref{tab:comparison}).
This additional power consumption results primarily from the resistive hybrid, which must simultaneously provide the DC bias required by the LVDS receiver and terminate the low-impedance transmission line.
Consequently, a moderate DC current flows through the hybrid continuously, consuming power that is not directly associated with signal generation.
This represents a trade-off between power efficiency and circuit simplicity.

\section{Loopback Function}

During the development and optimization of the transceiver, an additional function of the circuit was identified. The effect was first predicted by SPICE simulation and subsequently confirmed with the physical implementation.
Without any additional components, the transceiver inherently provides a loopback function: when no cable is connected to the coaxial port, the transmitted signal is received again by the same transceiver. The same effect occurs when a cable is connected but left open at its far end. The transmitted signal is then reflected at the open circuit and returns to the transceiver after a delay corresponding to twice the propagation time of the cable.

This behavior follows directly from the physics of the transmission line. An open-circuit termination reflects an incoming voltage wave with the same polarity and, for an ideal lossless line, the same amplitude. Because the resistive hybrid is matched to the line impedance, the effect is the same for any cable length, including a length of zero. From the viewpoint of the transceiver, the resulting received signal is therefore indistinguishable from a signal transmitted by an identical remote transceiver.

The loopback function was observed by applying a TTL-level test pattern from a signal generator and monitoring the returned signal with an oscilloscope. It was subsequently demonstrated using a USB-to-UART adapter configured for \SI{115200}{baud} and connected to a single transceiver with an open-ended coaxial cable. The transceiver behaved as a transparent loopback device, returning the transmitted UART data to the adapter. The loopback operation ceased when the far end of the cable was terminated with a \SI{50}{\ohm} load or connected to an idle second transceiver.

This property can be exploited for simple link-presence detection without additional hardware. A microcontroller or FPGA connected to the transceiver can transmit a suitable test pattern and compare the received data with the expected loopback response. Detection of the transmitted pattern suggests that no remote transceiver is connected. This feature may be useful for hot-pluggable point-to-point interfaces in which connecting or disconnecting a remote device is intended to trigger a specific system action.

Beyond its use as a communication interface, the same principle allows a single transceiver and an open-ended coaxial cable to be used as a logic-signal delay element. Such logic delay devices remain useful in nuclear and particle-physics detector systems, for example for delaying trigger or veto signals with nanosecond precision. Compared with using the cable directly as a delay line, the proposed circuit exploits the reflection at the open cable end and therefore provides a delay corresponding to twice the one-way propagation time of the cable for the same cable length.

\section{Conclusion}

A low-complexity passive resistive hybrid has been demonstrated as a means of realizing simultaneous bidirectional transmission of baseband logic signals over a single coaxial cable. The proposed transceiver combines a resistive hybrid implemented with nine standard SMD resistors, a single CMOS logic gate as line driver, and a standard LVDS receiver. No modulation or demodulation, transformers, adaptive balancing, or active analog or digital echo cancellation are required. The interface is protocol-independent and operates directly on asynchronous logic signals, including continuous logic-high and logic-low states. An analytical treatment of the hybrid network was used to determine the optimum resistor values for maximizing the differential signal amplitude at the receiver.

The approach deliberately exploits the controlled electrical characteristics of short to medium-length laboratory coaxial cables rather than attempting to maximize bandwidth over long or electrically challenging copper links. SPICE simulations show that the finite directional isolation of the resistive hybrid introduces deterministic edge timing errors, while preserving sufficient signal integrity for digital operation. For cable of type RG-178 and lengths below approximately \SI{6}{\meter}, the measured jitter agrees well with the simulated behavior and remains below approximately \SI{1}{\nano\second} peak-to-peak. For longer cables, an additional timing error is observed, which is attributed to frequency-dependent cable losses and the resulting degradation of signal edges. Nevertheless, the measured peak-to-peak timing error remains approximately below \SI{1.2}{\nano\second} for cable lengths up to \SI{10}{\meter}.

The intended data-rate target of \SI{200}{\mega\bit\per\second} was deliberately exceeded in an experiment using two independent randomized data streams at \SI{250}{\mega\bit\per\second}.
The circuit principle was also successfully adapted from \SI{50}{\ohm} to \SI{75}{\ohm} transmission lines by scaling the hybrid resistor values by a factor of 1.5. Furthermore, the transceiver exhibits an inherent loopback function when the coaxial port is left open, providing an additional possibility for link-presence detection without additional circuitry.

The proposed approach is intended primarily for laboratory and detector instrumentation, particularly where existing coaxial infrastructure is available and cable routing or vacuum feedthrough density is a concern. By allowing two independent signals to share a single coaxial line, the transceiver can reduce the required number of cables and feedthroughs. The absence of galvanic isolation limits its applicability to electrically compact setups in which ground-potential differences are small. Thus, rather than competing with high-speed communication systems optimized for long-distance transmission, the proposed transceiver deliberately trades maximum bandwidth and cable length for low circuit complexity, simple PCB integration, and direct compatibility with asynchronous logic signals.

%
%
%
%
%
%
%

\section{Methods}
\label{sec:methods}

Schematic capture and printed circuit board layout were performed using KiCad 9.0.6~\cite{kicad2025}. Circuit simulations were carried out using LTspice XVII~\cite{ltspice_17_0_30}. Simulation traces were imported and processed using the PyLTSpice~\cite{pyltspice} python library, with numerical analysis performed using NumPy and visualization using Matplotlib.

Periodic square-wave test signals were generated using a Tektronix AFG31252 arbitrary waveform generator, featuring an analog bandwidth of \SI{250}{\mega\hertz} and a sampling rate of \SI{2}{\giga Sa/\second}. Randomized serial data patterns were generated using a Keysight M8195A arbitrary waveform generator with an analog bandwidth of \SI{25}{\giga\hertz} and a sampling rate of \SI{65}{\giga Sa/\second}. The data pattern was the built-in pseudorandom binary sequence (PRBS) of length $2^9=512$ bits.

Waveforms were recorded using a LeCroy WaveRunner 620Zi digital oscilloscope with an analog bandwidth of \SI{2}{\giga\hertz} at a sampling rate of \SI{10}{\giga Sa/\second}. Where required, differential voltage waveforms and edge steepness were measured using a LeCroy ZD1500 active differential probe, specified for an analog bandwidth of \SI{1.5}{\giga\hertz}, a differential input resistance of \SI{120}{\kilo\ohm}, and an input capacitance of \SI{1}{\pico\farad}.

\section*{Acknowledgements}
The author thanks his colleagues from the GSI Experiment Electronics group for numerous stimulating discussions. In particular, the author acknowledges Dawid Madzelan for raising questions regarding directional couplers that helped initiate this work.

The author used an AI-based language model for assistance in language refinement and manuscript organization. All technical content, analysis, and conclusions are the author’s own.

The publication is funded by the Open Access Publishing Fund of GSI Helmholtzzentrum fuer Schwerionenforschung.

\section*{Conflict of Interest}

The authors have no conflicts to disclose.

\section*{Data Avaliability}
The data that support the findings of this study are available from the corresponding author upon reasonable request.

\bibliography{bib_legacy}

\end{document}